\documentclass[prb,reprint,twocolumn,superscriptaddresss]{revtex4-1}
\def\be{\begin{equation}}
\def\ee{\end{equation}}
\def\bea{\begin{eqnarray}}
\def\eea{\end{eqnarray}}
\def\bi{\begin{itemize}}
\def\ei{\end{itemize}}
\usepackage{braket}
\def\nb{\nonumber}
\usepackage{xcolor}
\usepackage{graphicx}
\usepackage{amsmath,amstext,amssymb,bm,color,times}
\usepackage[english]{babel}
\usepackage[T1]{fontenc}
\usepackage[utf8]{inputenc}
\usepackage[colorlinks=true,citecolor=blue,linkcolor=magenta]{hyperref}

\begin{document}

\title{Inhomogeneity induced shortcut to adiabaticity in Ising chains with long-range interactions} 

\author{Aritra Sinha} 
\affiliation{Jagiellonian University, Institute of Theoretical Physics, {\L}ojasiewicza 11, PL-30348 Krak\'ow, Poland}
\author{Debasis Sadhukhan} 
\affiliation{Jagiellonian University, Institute of Theoretical Physics, {\L}ojasiewicza 11, PL-30348 Krak\'ow, Poland}
\author{Marek M. Rams}
\affiliation{Jagiellonian University, Institute of Theoretical Physics, {\L}ojasiewicza 11, PL-30348 Krak\'ow, Poland}
\author{Jacek Dziarmaga}
\affiliation{Jagiellonian University, Institute of Theoretical Physics, {\L}ojasiewicza 11, PL-30348 Krak\'ow, Poland}
\date{\today}

\begin{abstract}
Driving a homogeneous system across a quantum phase transition in a quench-time $\tau_Q$ generates excitations on wavelengths longer than the Kibble-Zurek (KZ) length $\hat\xi\propto\tau_Q^{\nu/(1+z\nu)}$ within the KZ time window $\hat t\propto\tau_Q^{z\nu/(1+z\nu)}$, where $z$ and $\nu$ are the critical exponents. Quenches designed with local time-dependent inhomogeneity can introduce a gap in the spectrum. They can be parametrized by a time- and space-dependent distance $\epsilon$ from the critical point in the parameter space of the Hamiltonian: $\epsilon(t,x)=\frac{t-x/v}{\tau_Q} \equiv \theta\cdot (v t-x). $ For a variety of setups with short-range interactions, they have been shown to suppress excitations if the spatial velocity $v$ of the inhomogenous front is below the characteristic KZ velocity $\hat v \propto \hat\xi/\hat t$. Ising-like models with long-range interactions can have no sonic horizon, spreading information instantaneously across the system. Usually, this should imply that inhomogenous transitions will render the dynamics adiabatic regardless of the velocity of the front. However, we show that we get an adiabatic transition with no defects only when the inhomogeneous front moves slower than the characteristic crossover velocity $\tilde v \propto \theta^{(z-1)\nu/(1+\nu)}$, where $\theta$ is the slope of the inhomogeneous front at the critical point. The existence of this crossover velocity and adiabaticity of the model results from the energy gap in the quasiparticle spectrum that is opened by the inhomogeneity. This effect can be employed for efficient adiabatic quantum state preparation in systems with long-range interactions.
\end{abstract}
\maketitle

\section{ Introduction } 

The Kibble-Zurek mechanism (KZM) originated from a scenario for defect creation in cosmological symmetry-breaking phase transitions~\cite{K-a, *K-b, *K-c}. As the Universe cools, causally disconnected regions must choose broken symmetry vacua independently. Such random choices lead to topologically nontrivial configurations that survive as topological defects. In the cosmological setting, the average size of the causally connected regions---and the average density of defects---is set by the Hubble radius at the time of the transition. This early Universe scenario relies on the speed of light and does not apply to the laboratory phase transitions. For the latter, the dynamical theory \cite{Z-a, *Z-b, *Z-c, Z-d} employs critical exponents of the transition and the quench time to predict the scaling of the resulting density of defects. The KZM was successfully tested using numerical simulations \cite{KZnum-a,KZnum-b,KZnum-c,KZnum-d,KZnum-e,KZnum-f,KZnum-g,*KZnum-h,*KZnum-i,KZnum-j,KZnum-k,KZnum-l,KZnum-m,that} and laboratory experiments in condensed matter systems \cite{KZexp-a,KZexp-b,KZexp-c,KZexp-d,KZexp-e,KZexp-f,KZexp-g,KZexp-gg,KZexp-h,KZexp-i,KZexp-j,KZexp-k,KZexp-l,KZexp-m,KZexp-n,KZexp-o,KZexp-p,KZexp-q,KZexp-r,KZexp-s,KZexp-t,KZexp-u,KZexp-v,KZexp-w,KZexp-x}. More recently, the KZM was adapted to quantum phase transitions \cite{QKZ1,QKZ2,QKZ3,d2005,d2010-a, d2010-b}. Theoretical developments \cite{QKZteor-a,QKZteor-b,QKZteor-c,QKZteor-d,QKZteor-e,QKZteor-f,QKZteor-g,QKZteor-h,QKZteor-i,QKZteor-j,QKZteor-k,QKZteor-l,QKZteor-m,QKZteor-n,QKZteor-o,KZLR1,KZLR2,KZLR3,QKZteor-q,QKZteor-r,QKZteor-s,QKZteor-t,sonic,QKZteor-u,QKZteor-v,QKZteor-w,QKZteor-x} and experimental tests \cite{KZexp-gg, QKZexp-b, QKZexp-c, QKZexp-d, QKZexp-e, QKZexp-f, QKZexp-g,deMarco2,Lukin18,adolfodwave} of quantum KZM (QKZM) followed. The recent experiment~\cite{Lukin18}, where a quantum Ising chain in the transverse field is emulated using Rydberg atoms, is consistent with the theoretically predicted scalings~\cite{QKZ2,QKZ3,d2005}. 

In a cartoon version, whose limitations---but also basic correctness---have been discussed in Ref.~\onlinecite{sonic}, the dynamics of the system literally freezes-out in the neighborhood of the critical point due to the critical slowing down/closing of the energy gap. In a QKZM a system initially prepared in its ground state is smoothly ramped across a critical point to the other side of the quantum phase transition. A distance from the critical point, measured by a dimensionless parameter $\epsilon$ in a Hamiltonian, can be linearized close to the critical point as
\begin{equation}
\epsilon(t)=\frac{t}{\tau_Q},
\label{epsilont}
\end{equation}
where $\tau_Q$ is a quench time. Initially, far from the critical point, the evolution is adiabatic and the system follows its adiabatic ground state. The adiabaticity fails at $-\hat t$ when the reaction time of the system, given by the inverse of the gap, becomes slower than the timescale $|\epsilon/\dot \epsilon| = |t|$ on which the transition is being imposed. The gap closes like $\Delta\propto|\epsilon|^{z\nu}$, where $z$ and $\nu$ are the dynamical and correlation length exponents, respectively. From the equation $|t|\propto |t/\tau_Q|^{-z\nu}$ we obtain $\hat t\propto \tau_Q^{z\nu/(1+z\nu)}$ and the corresponding $\hat\epsilon=\hat t/\tau_Q\propto\tau_Q^{-1/(1+z\nu)}$. In the cartoon ``freeze-out'' version of the impulse approximation the ground state at $-\hat\epsilon$, with a corresponding correlation length,
\begin{equation}
\hat\xi \propto \tau_Q^{\nu/(1+z\nu)},  
\label{hatxi}
\end{equation}
is expected to survive until $+\hat t$, when the evolution can restart. In this way, $\hat\xi$ becomes imprinted on the initial state for the final adiabatic stage of the evolution after $+\hat t$. Oversimplified as it is, the adiabatic-impulse-adiabatic approximation predicts correct scaling of the characteristic lengthscale with $\tau_Q$, see Eq.~\eqref{hatxi}, and the timescale 
\be 
\hat t \propto \hat\xi^z.
\label{hatt}
\ee 
The post-quench density of excitations is determined by $\hat\xi$ within this scenario.  

\begin{figure}
    \centering
    \includegraphics[width=0.99\columnwidth]{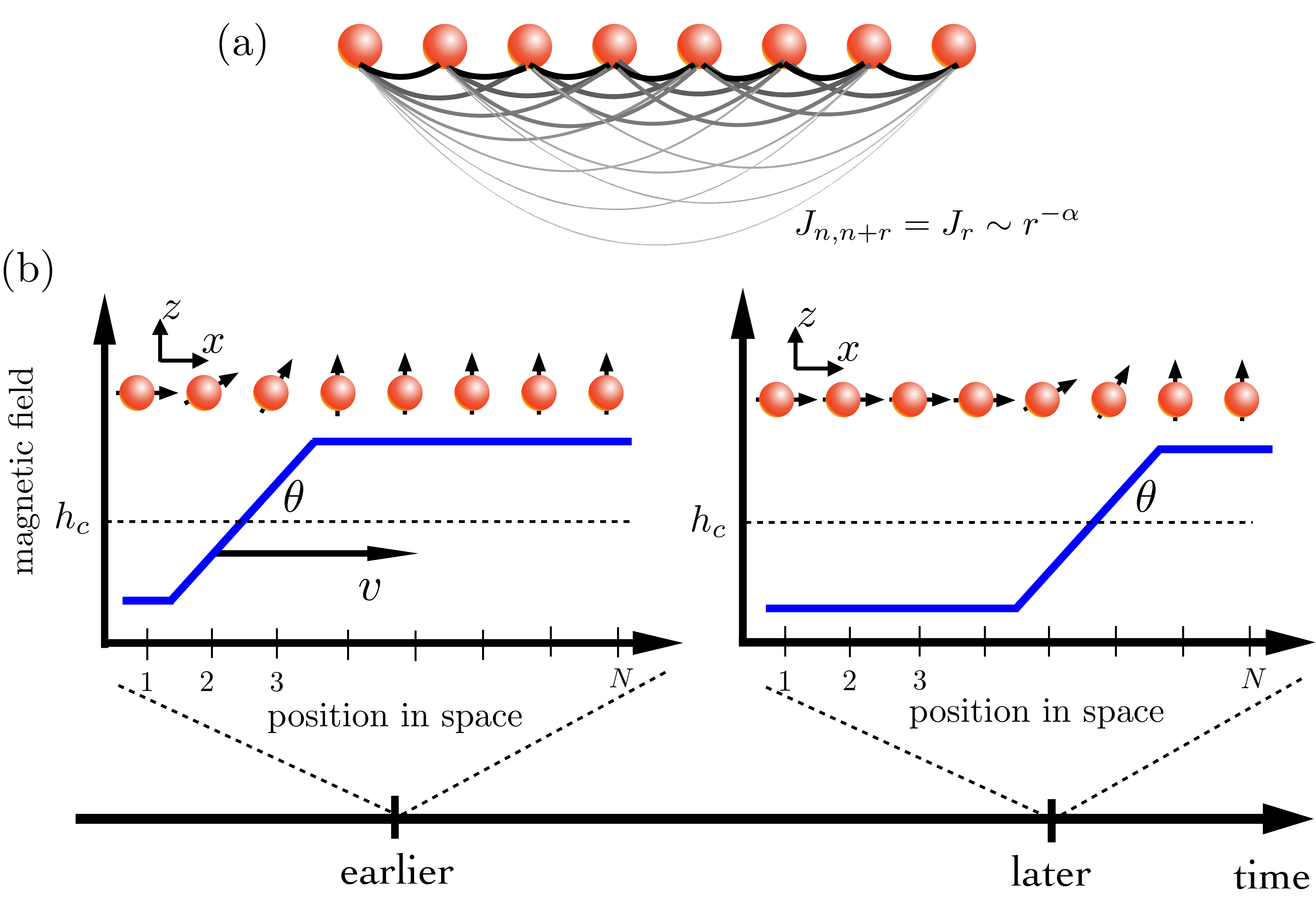}
    \caption{{\it Schematic illustration of an inhomogeneous driving protocol in a system with long-range interactions.} In (a), we indicate the long-range couplings considered in this article, vanishing as a power-law with distance. In (b), the magnetic field is position-dependent and interpolates in space between disordered and ordered phases. The inhomogeneous front has a linear slope equal to $\theta$ and sweeps across the system with a spatial velocity $v$. The interplay between the velocity, slope, and universal critical exponents of the model paves the way for enhancing the adiabaticity of the transition across the critical points.}
    \label{fig:tanh}
\end{figure}

Furthermore, $\hat\xi$ and $\hat t$ can be combined into the KZ spatial-velocity scale 
\be 
\hat v \propto \frac{\hat\xi}{\hat t} \propto \tau_Q^{(1-z)\nu/(1+z\nu)}.
\label{hatv}
\ee 
In homogeneous systems with short-range interactions this is the maximal velocity of quasiparticles excited by the quench~\cite{sonic}. This is why this speed limit is believed to be central for the shortcuts to adiabaticity via an inhomogeneous KZM. Therein, the external driving field has a smooth position dependence, characterized by
\begin{equation}
\epsilon(t,x)=\frac{t-x/v}{\tau_Q} \equiv \theta v t- \theta x.  
\label{epsilontx}
\end{equation}
Here $\theta=(v \tau_Q)^{-1}$ is a slope of the time-dependent ramp and $x=v t$ is the location where the driving field assumes the critical value. The ramp is gradually taking the system across the critical point one part after another, see Fig.~\ref{fig:tanh}. In systems with short-range interactions, the velocity of the driven critical front $v$ that is below $\hat v$ was shown to pave the way to adiabatic dynamics, both for classical\cite{ inhomo_classical-a, KZnum-c, inhomo_classical-c, inhomo_classical-d, inhomo_classical-e, inhomo_classical-f} and quantum\cite{inhomo_quantum-a, *inhomo_quantum-aa, inhomo_quantum-b, *inhomo_quantum-c, inhomo_quantum-d, inhomo_quantum-e, inhomo_quantum-f, inhomo_quantum-h} systems. An explanation appealing to causality is that the initial choice of symmetry breaking to the left of the critical point (i.e., in the part of the system that has crossed the critical point earlier, see Fig.~\ref{fig:tanh}) is communicated across it, and biases the parts of the system that are being taken across the critical point to make the same choice. The bias is not feasible in the supersonic regime, $v\gg\hat v$, where the usual homogeneous KZM with quench time $\tau_Q$ is ultimately recovered.

The homogeneous QKZM in systems with genuinely long-range interactions was considered in recent publications \cite{KZLR1, KZLR2, sonic}. As no sonic horizon is observed for sufficiently long-ranged interactions \cite{LRcone1,LRcone2,LRcone3,LRcone4} the causality argument becomes problematic. Indeed, at least in exactly solvable models of this class, the dynamical exponent $z<1$ implies (via quasiparticle dispersion $\omega\propto k^z$) that quasiparticle group velocity, $d\omega/dk\propto k^{z-1}$, diverges when $k\to0$. The softest quasiparticles, which are most excited by the quench, have infinite propagation velocity. Furthermore, given that the QKZM excites quasimomenta up to $\hat k\propto\hat{\xi}^{-1}$, whose group velocity is $\hat k^{z-1}\propto\hat{\xi}/\hat t\propto\hat v$, the KZ velocity is not an upper but a lower speed limit. Therefore, there should be no speed limit (in the usual sense of this word) to suppress communication across the critical point when its velocity $v\gg\hat v$. The usual adiabatic evolution is expected for $v\ll\hat v$, but, based on causality alone, adiabaticity in the regime $v\gg\hat v$ cannot be excluded. Surprisingly, we observe a clear-cut crossover between an adiabatic regime and effectively homogeneous KZM regime near $\hat v$, challenging the conventional notions of causality. 

A possible clue to the explanation may lay in a variant of the  KZM argument where the transition happens in space \cite{Platini_2007, Dorner_KZspace_2008, Collura_2009, Damski_2009,  inhomo_quantum-a, Lacki_KZspace_2017} rather than time. Here we follow the discussion in Ref.~\onlinecite{inhomo_quantum-a} where it was thoroughly illustrated in the standard short-ranged quantum Ising chain. In a similar way as in Eq. \eqref{epsilont}, we make a linear approximation near the critical point:
\begin{equation}
\epsilon(x)=-\theta x.  
\label{epsilonx}
\end{equation}
In the first ``local density approximation'' we would expect the order parameter to behave as if the system were locally uniform: it is non-zero only in the symmetry broken phase, for $x<0$, and tends to zero like $|\theta x|^\beta$ when $x$ approaches the critical point at $x=0^-$. However, this local approximation leads to a contradiction because it predicts that a local correlation length diverges like $\xi(x)\propto|\theta x|^{-\nu}$ as the critical point at $x=0$ is approached. As the diverging $\xi$ is the shortest healing length on which the order parameter can adapt to changing $\epsilon(x)$, the local approximation must fail when the distance remaining to the critical point, $|x|$, becomes comparable to the local healing length proportional to $|\epsilon(x)|^{-\nu}$. It breaks down at a distance $|x|\propto\tilde\xi$ from the critical point, where
\begin{equation}
    \tilde{\xi} \propto \theta^{-\nu/(1+\nu)}.
    \label{tildexi}
\end{equation}
From $x\propto -\tilde{\xi}$ the evolution of the order parameter with $x$ becomes ``impulse'', i.e., the order parameter does not change until $x\propto +\tilde\xi$ on the symmetric side of the transition where it begins to follow the local $\epsilon(x)$ again. This way, the QKZM in space predicts that the order parameter penetrates the symmetric phase to a depth proportional to $\tilde\xi$. In other words, the non-analytic critical point is rounded off on the length scale $\tilde\xi$. On this basis, in the spirit of the critical scaling hypothesis, we expect a non-zero energy gap scaling as
\begin{equation}
    \tilde \Delta \propto \tilde\xi^{-z} \propto \theta^{z\nu/(1+\nu)}.
    \label{tiledDelta}
\end{equation}
The local density approximation would predict a vanishing gap for quasiparticles localized at the critical point, but, as the localization is at odds with the zero-gap, the energy gap becomes non-zero and is set by the slope $\theta$ of the critical front.

The quasiparticles with this minimal gap are localized within distance $\tilde\xi$ from the critical point. This is why, when the front is made to move such as Eq.~\eqref{epsilontx}, the relevant transition rate is set by a ratio $v/\tilde\xi$. A comparison between this rate and the gap $\tilde\Delta$ yields a crossover velocity between an adiabatic and an effectively-homogeneous transition,
\begin{equation}
    \tilde v \propto \tilde\xi \tilde\Delta \propto \theta^{(z-1)\nu/(1+\nu)}.
    \label{tildev}
\end{equation}
For $v\gg\tilde v$, we expect $\tilde v$---and the inhomogeneity as such, from which it originates---to be irrelevant and the transition to proceed as if were effectively homogeneous. On the other hand, for $v\ll\tilde v$, excitations should be adiabatically suppressed by the finite quasiparticle gap $\propto\tilde\Delta$. 

As a final remark, notice that the velocity scale $\tilde v$ is not unrelated to $\hat v$ in Eq.~\eqref{hatv}. When $\theta=1/v\tau_Q$ is inserted into Eq.~\eqref{tildev} and solved with respect to $v$ we obtain $\hat v$ in Eq.~\eqref{hatv}. However, the origin of $\tilde v$ is the finite gap induced by the inhomogeneity and not the causality that becomes problematic for long-range interactions.

This hypothesis of adiabatic-homogeneous crossover at $\tilde v$ brings us close to the concept of shortcuts to adiabaticity~\cite{TORRONTEGUI2013117,del_Campo_2019, Odelin2019}. They are a set of fast processes leading one to the desired final result, which otherwise would be obtained by changing the parameters of the system adiabatically slowly. Quenches with local time-dependent spatial inhomogeneity had already been established as tools for bringing about a shortcut to adiabaticity~~\cite{inhomo_quantum-a, *inhomo_quantum-aa, inhomo_quantum-b, *inhomo_quantum-c, inhomo_quantum-d, inhomo_quantum-e, inhomo_quantum-f, inhomo_quantum-g, inhomo_quantum-h,inhomo_quantum-i,inhomo_quantum-j,inhomo_quantum-k}. For homogeneous phase transitions, the evolution becomes trivially adiabatic when the characteristic KZ length scale of defects $\hat{\xi}$, given by Eq.~\eqref{hatxi}, becomes longer than the system size $N$ for a particular quench time $\tau_{Q}^{\rm adiab}$. One can show that an inhomogeneous transition with an adequately selected spatial slope can allow one to achieve adiabaticity on shorter timescales than the usual homogeneous quench timescales $\tau_{Q}^{\rm adiab}$, thus achieving the promised shortcut to adiabaticity.

Having thus outlined theoretical foundations, in the following, we illustrate the above general considerations with two examples: the long-range extended Ising model in Sec.~\ref{sec:LR} and the long-range Ising model in Sec.~\ref{sec:LRI}. The former one is equivalent, via the Jordan-Wigner transformation, to the long-range Kitaev model. It is solvable by the Bogoliubov transformation and, therefore, provides analytic insights into the problem. In contrast, the latter requires extensive matrix product state simulations but appears more realistic from the experimental point of view. With the theory substantiated by the examples, in Sec.~\ref{sec:shortcuts}, we discuss its practical implications for shortcuts to adiabaticity. Finally, we summarize in Sec.~\ref{sec:summary}.  

\section{ Long-range extended Ising chain } 
\label{sec:LR}

The long-range extended Ising chain reads
\bea
&& 
H =
-\sum_{n=1}^N
\left(
 h_n \sigma^z_n +
\sum_{r=1}^{N-n} 
J_r ~ \sigma^x_n ~ Z_{n,r} ~ \sigma^x_{n+r}
\right),
\label{Hamil_LR} \\
&&
Z_{n,r} = 
\left\{
\begin{array}{cc}
\prod_{m=n+1}^{n+r-1} \sigma^z_m & {\rm ,~~ when ~~} r>1, \\
1                                & {\rm ,~~ when ~~} r=1, 
\end{array}
\right.
\nonumber 
\eea
Here $h_n$ is the strength of the transverse magnetic field, and $J_r$ is the interaction strength between any pair of spins separated by a distance $r$. With the string operator $Z_{n,r}$ between the spins the model is equivalent to the long-range Kitaev model~\cite{LRKitaev1} in Eq. (\ref{HcLR}). In this work, we will consider power-law long-range interaction:
\be
 J_r = {\cal N}_\alpha~ \frac{1}{r^\alpha},
\ee
where ${\cal N}_\alpha$ is a normalization constant. For $\alpha>1$ it is possible to normalize $\sum_{r} J_r=1$ with ${\cal N}_\alpha$ expressed by the Riemann zeta function, ${\cal N}_\alpha=1/\zeta(\alpha)$. 

In the homogeneous case, we set the transverse field strength, 
\be 
h_n=h_c-\epsilon,
\label{epsilon}
\ee
where $\epsilon = 0$ corresponds to the quantum critical point at $h_c=1$ between the paramagnetic ($\epsilon<0$) and ferromagnetic ($\epsilon>0$) phases of the system in the thermodynamic limit. 
 
Thanks to the string operators in the long-range interaction terms, the model can be mapped to a quadratic free-fermionic problem. After the Jordan-Wigner transformation,
\begin{eqnarray}
\sigma^x_n &=&\left( c_n + c_n^\dagger \right)
 \prod_{m<n}(1-2 c^\dagger_m c_m)~, \nonumber \\
\sigma^z_n &=& 1~-~2 c^\dagger_n  c_n~ \label{JordanWigner}, 
\end{eqnarray}
introducing fermionic annihilation operators $c_n$, the Hamiltonian \eqref{Hamil_LR} becomes  
\begin{eqnarray}
H = \sum_{n=1}^N \Big[&&2h_n c_n^\dag c_n  \nb \\
  &&- \sum_{r=1}^{N-n} J_r\left( c_n^\dag c_{n+r} + c_n^\dag c_{n+r}^\dag+{\rm H.c.}\right) \Big],
\label{HcLR}
\end{eqnarray}
up to a constant additive term. This Hamiltonian is also known as a long-range Kitaev chain~\cite{LRKitaev1}. It can be further reshaped as 
\begin{eqnarray}
H = \sum_{n,m=1}^N c_n^\dagger {A}_{n,m} c_m+\frac{1}{2} \sum_{n,m=1}^N \Big(c_n^\dagger {B}_{n,m} c_{m}^\dagger+ \mbox{H.c.}\Big),
\label{HAB}
\end{eqnarray}
where 
\begin{eqnarray}
A_{n,n} &=& 2h_n, ~~ A_{n,m=n+r} = -J_r = A_{m=n+r,n}\\
B_{n,n} &=& 0, ~~ B_{n,m=n+r} = J_r = -B_{m=n+r,n}
\end{eqnarray}
for open boundary conditions.

The quadratic Hamiltonian can be diagonalized by a Bogoliubov transformation to new fermionic Bogoliubov quasiparticle annihilation operators, 
\be
\gamma_m = \sum_{n=1}^N \left(u_{nm}^* c_n + v_{nm} c_n^\dagger\right),
\label{nFB}
\ee
where the $N$-dimensional vectors ${\bf u}_m$ and ${\bf v}_m$, for $m=1,\ldots,N$, satisfy the stationary Bogoliubov-de Gennes equations: 
\begin{eqnarray}
  A \cdot {\bf u}_m + B \cdot {\bf v}_m & = & \omega_m {\bf u}_m, \nonumber\\
 -B \cdot {\bf u}_m - A \cdot {\bf v}_m & = & \omega_m {\bf v}_m. 
 \label{sBdG}
 \end{eqnarray}
Unlike the homogeneous case with periodic boundary conditions, for which Eq.~\eqref{sBdG} can be reduced by a Fourier transform to independent $2\times 2$ blocks, we need to diagonalize the full $2N\times 2N$ matrix numerically. The spectrum of Eq.~\eqref{sBdG} is given by $\pm \omega_m$, where we fix $\omega_m\ge 0$. Following transformation \eqref{nFB}, the Hamiltonian in Eq.~\eqref{HAB} becomes 
\begin{eqnarray}
H = \sum_{m=1}^N \omega_m \left(\gamma_m^\dagger \gamma_m - \frac12\right).
\label{HkLR}
\end{eqnarray}
Here, the Bogoliubov quasiparticles \eqref{nFB} are defined by the eigenmodes 
$({\bf u}_m,{\bf v}_m)$ with the positive eigenvalues $+\omega_m$. Correspondingly, $({\bf u}_m^-,{\bf v}_m^-)$ are the negative eigenmodes with eigenvalues $-\omega_m$ which define the Bogoliebov quasiparticles $\gamma_{m}^{-}=\sum_n({u}_{nm}^{-})^{*}c_{n}+v_{nm}^{-}c_{n}^{\dag}$. 

The Hamiltonian commutes with the parity operator 
\be 
{\cal P} = 
\prod_{n=1}^N\sigma_n^z = 
\prod_{n=1}^N \left(1-2c_n^\dag c_n\right). 
\label{eq:parity}
\ee 
In the following, we consider transitions that begin in the ground state with positive parity. As the parity is conserved during the unitary evolution, we can confine ourselves to the subspace of states with an even number of Bogoliubov quasiparticles.

In the thermodynamic limit, $N\to\infty$, the homogeneous chain would have zero energy gap at the critical point at $\epsilon=0$. There is a finite gap in a finite chain that goes to zero as $N^{-z}$ when $N\to\infty$. Here $z=\alpha-1$ is the dynamical critical exponent that governs the quasiparticle dispersion, $\omega_k \propto k^z$,  where $k$ is the pseudo-momentum. The correlation length exponent is $\nu=1/(\alpha-1)$.
Notice that for $\alpha<2$ we have $z<1$ and the quasiparticle group velocity, 
\be 
v_k=\frac{d\omega_k}{dk}\propto k^{z-1}, 
\ee 
diverges for $k\to 0$. As the softest modes with small $k$ are most excited in typical quench protocols, there is no sonic horizon for the propagation of the soft excitations, as might have been expected in a model with long-range interactions.  

When we ramp down the parameter $\epsilon$ as a linear function of time, given by Eq.~\eqref{epsilont}, driving the Hamiltonian from the initially prepared paramagnetic ground state to the ordered phase, the resulting state ends up being \emph{defective} as compared to the instantaneous ground state at the end of the evolution. These defects are quantified by the density of excited quasiparticles that can be calculated at each time by projecting the time-dependent Bogoliubov modes  onto the corresponding instantaneous static negative Bogoliubov modes 
\begin{eqnarray}
\centering
d_{ex}(t) = \frac{1}{N} \sum_{m}^N\sum_{n}^N \left|\langle {\bf u}_{m}^{{-}},{\bf v}_{m}^{{-}}|{\bf u}_{n}(t),\textbf{v}_{n}(t)\rangle\right|^2.
\label{doe}
\end{eqnarray}
The time-dependent modes, $({\bf u}_{n}(t),\textbf{v}_{n}(t))$, follow in the Heisenberg picture from the time-dependent variant of the Bogoliubov-de Gennes equations~\eqref{sBdG}, see, e.g., Ref.~\onlinecite{d2010-a}.

\begin{figure}[t!]
\includegraphics[width=0.99\columnwidth]{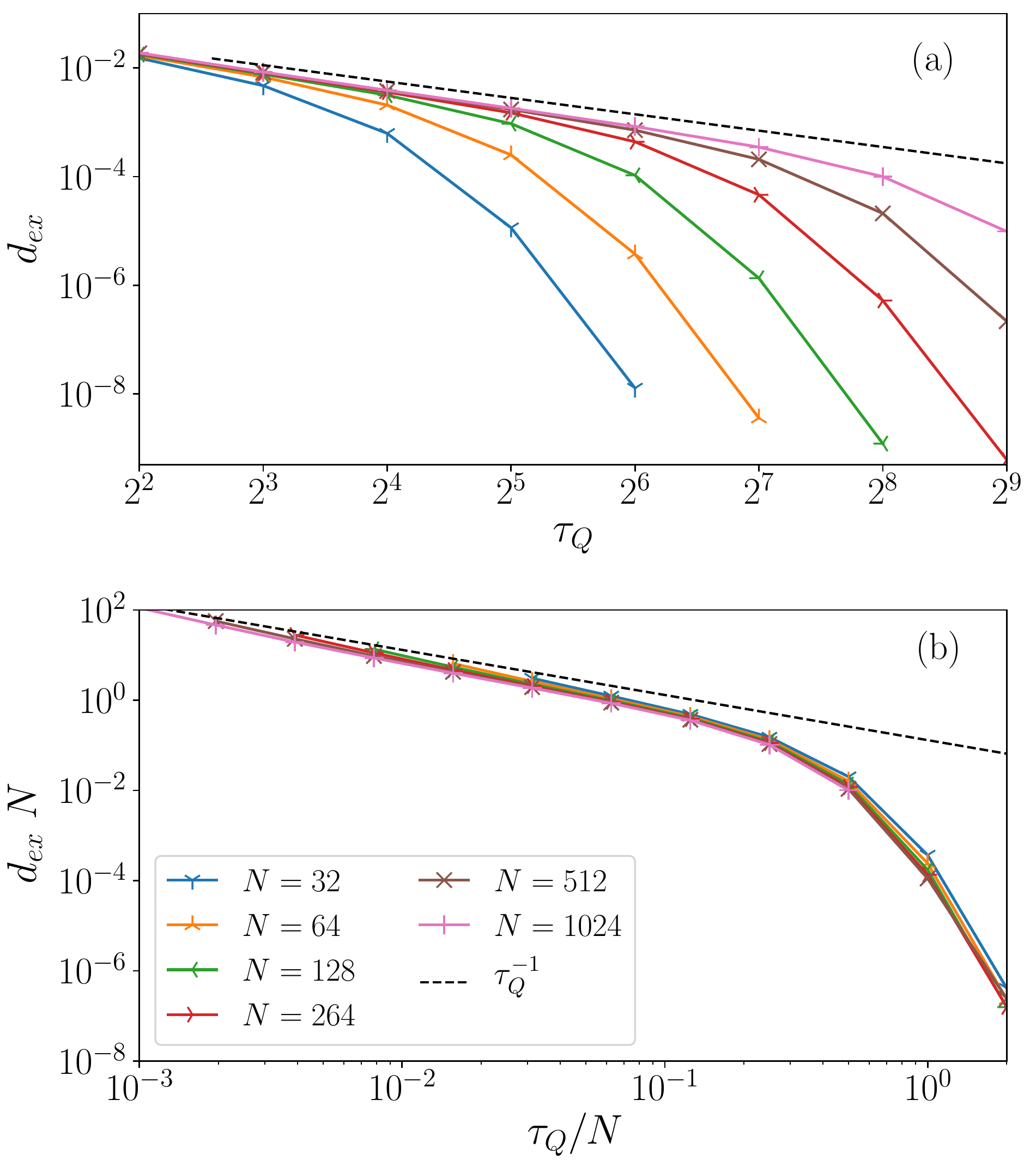}
\caption{{\it Density of excited quasiparticles at the end of the homogeneous quench in the long-range extended Ising model}. The ramp follows Eq.~\eqref{homoramp}, crossing the critical point at quench time $\tau_Q$. The long-range interactions vanish with exponent $\alpha=3/2$. In (a), the density decays like $\tau_Q^{-1/2(\alpha-1)}$ for faster quenches as predicted by Eq.~\eqref{dexhomo}. Finite-size effects become visible in the limit of larger $\tau_Q$. We show their universal character in (b), where we present rescaled density $d_{\rm ex}N$ in the function of rescaled quench time $\tau_Q/N^{2(\alpha-1)}$. The collapse of the plots for different system sizes $N$ demonstrates the crossover from the KZ to the adiabatic regime predicted by Eq.~\eqref{homoadiab}.
}
\label{fig:homo}
\end{figure}

\subsection{Homogeneous transition}
\label{sec:KZhomo}

After a homogeneous transition, we expect a finite density of excitations
\be 
d_{\rm ex} \propto \hat\xi^{-1} \propto \tau_Q^{-\nu/(1+z\nu)} = \tau_Q^{-1/2(\alpha-1)},
\label{dexhomo}
\ee 
with second equation for extended long-range Ising model considered here, and as long as KZ lengthscale $\hat\xi\ll N$ (but still large enough to be in the scaling limit). For slower quenches with
\be 
\tau_Q\gg N^{(1+z\nu)/\nu} =  N^{2(\alpha-1)},
\label{homoadiab}
\ee 
the density of quasiparticles excited when crossing the critical point should decay exponentially with $\tau_Q$, suppressed by a finite gap at the critical point.

In our numerical simulation we ramp the transverse field as
\begin{eqnarray}
h(t) = h_c - h_{c}\tanh\left(\frac{t}{h_{c}\tau_{Q}}\right),
\label{homoramp}
\end{eqnarray} 
from $h\approx 2$ at the initial time $t_i = -C\tau_Q h_{c}$ to $h\approx 0$ at the final time $t_f = C\tau_Q h_{c}$, with $C$ set to a large value.  The ramp is approximately linear, $h(t)\approx h_c - \frac{t}{\tau_Q}$, near the critical $h_c=1$ crossed at $t=0$, but additionally it reduces discontinuities in time-derivative at the initial and final times in order to suppress additional non-KZ excitations at the beginning of the ramp. We set $C = 3$ in the case of long-range extended Ising model which is large enough to obtain $C$-independent results. We collect the results of the numerical simulations in Fig.~\ref{fig:homo}.
 In this and the following figures, we show a representative  $\alpha=3/2$, though we have also explored other $1<\alpha<2$. For smaller $\alpha$, the results become somewhat obscured by the finite-size effects, and for larger $\alpha$, they systematically tend toward the effectively-local model for $\alpha\geq2$, consistently with the KZ scaling predictions.

\begin{figure}[t!]
\includegraphics[width=0.99\columnwidth]{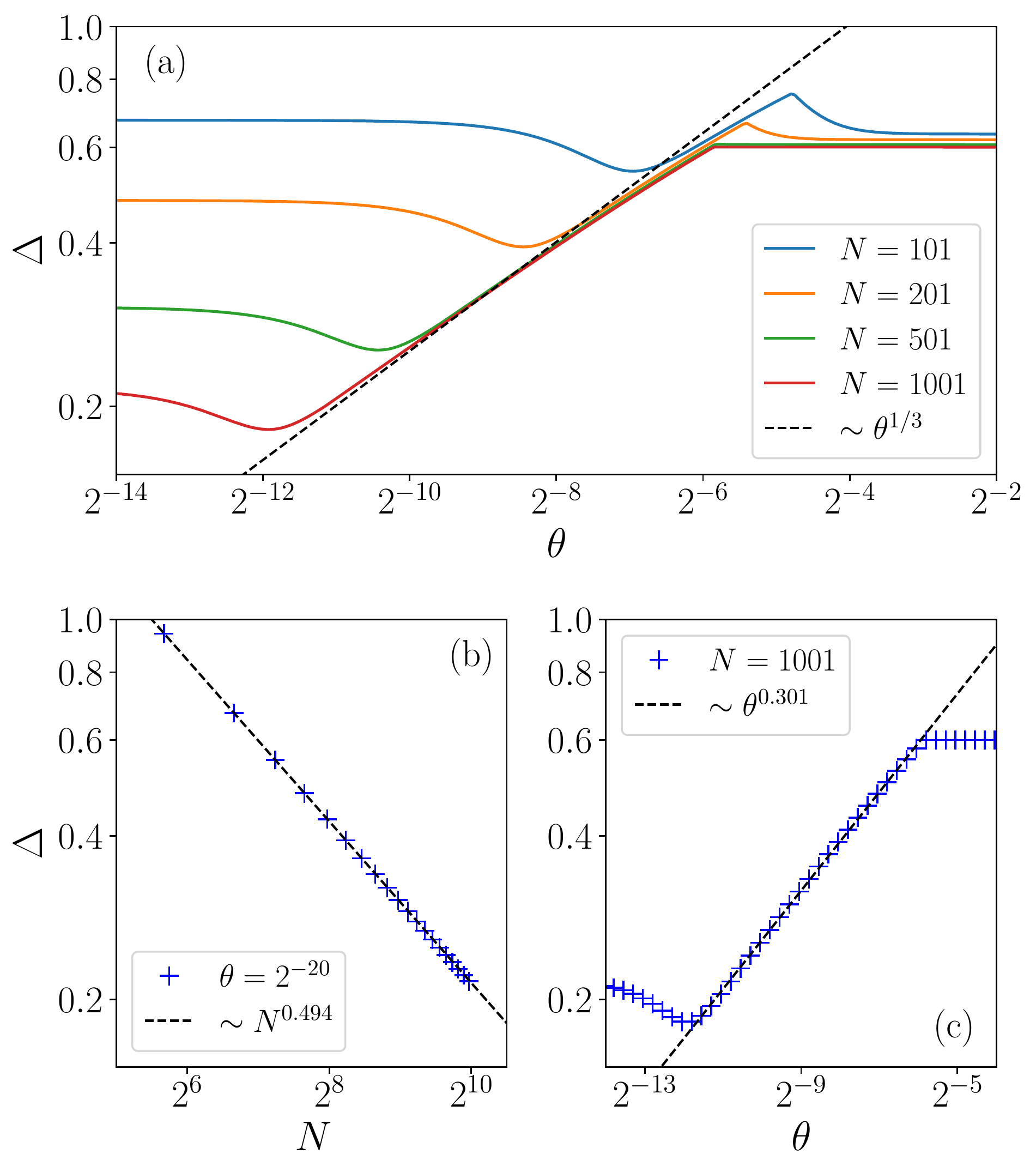}
\caption{\emph{The energy gap as a function of the inclination of the inhomogeneous front in the long-range extended Ising model.}
In (a),
the energy gap $\Delta=\omega_0+\omega_1$ for $\alpha=3/2$ as a function of the slope $\theta$ for different system sizes $N$. As predicted by Eq.~\eqref{hatDelta} the central part follows the scalling $\Delta\propto\theta^{1/3}$.
In (b), 
the gap for a very shallow---practically homogeneous---ramp with $\theta =2^{-20}$. The solid line is the best fit $\Delta = 6.58  ~N^{-0.494}$ which is consistent with $\Delta\propto N^{-z}=N^{-1/2}$ in a finite homogeneous system.
In (c),
zoom on the central part of panel (a) for $N=1001$. The dashed line is the best fit $\Delta = 2.07 ~ \theta^{0.301}$ which is close to the expected $\Delta=\hat\Delta\propto\theta^{1/3}$ predicted by Eq.~\eqref{hatDelta}.
}
\label{fig:gapinspace}
\end{figure}

\subsection{Static transition in space}
\label{sec:inspace}

Before the dynamics of an inhomogeneous transition, it is instructive first to study a transition in space~\cite{ Platini_2007,  Dorner_KZspace_2008, Damski_2009, Collura_2009, Lacki_KZspace_2017}. Towards this end, we consider a static inhomogeneous field,
\be
h_n = h_c + h_c \tanh\big((n-n_c)\theta/h_c\big),
\label{tanh}
\ee
that can be linearized near the critical point as $h_n \approx h_c+\theta \cdot (n-n_c)$, with the slope equal to $\theta$. Here $n_c$ is the location where the field reaches the critical value, $h_{n_c}=h_c$. The system is in the broken symmetry phase for $n < n_c$ and in the symmetric phase for $n>n_c$.

\begin{figure}[t]
\includegraphics[width=0.99\columnwidth]{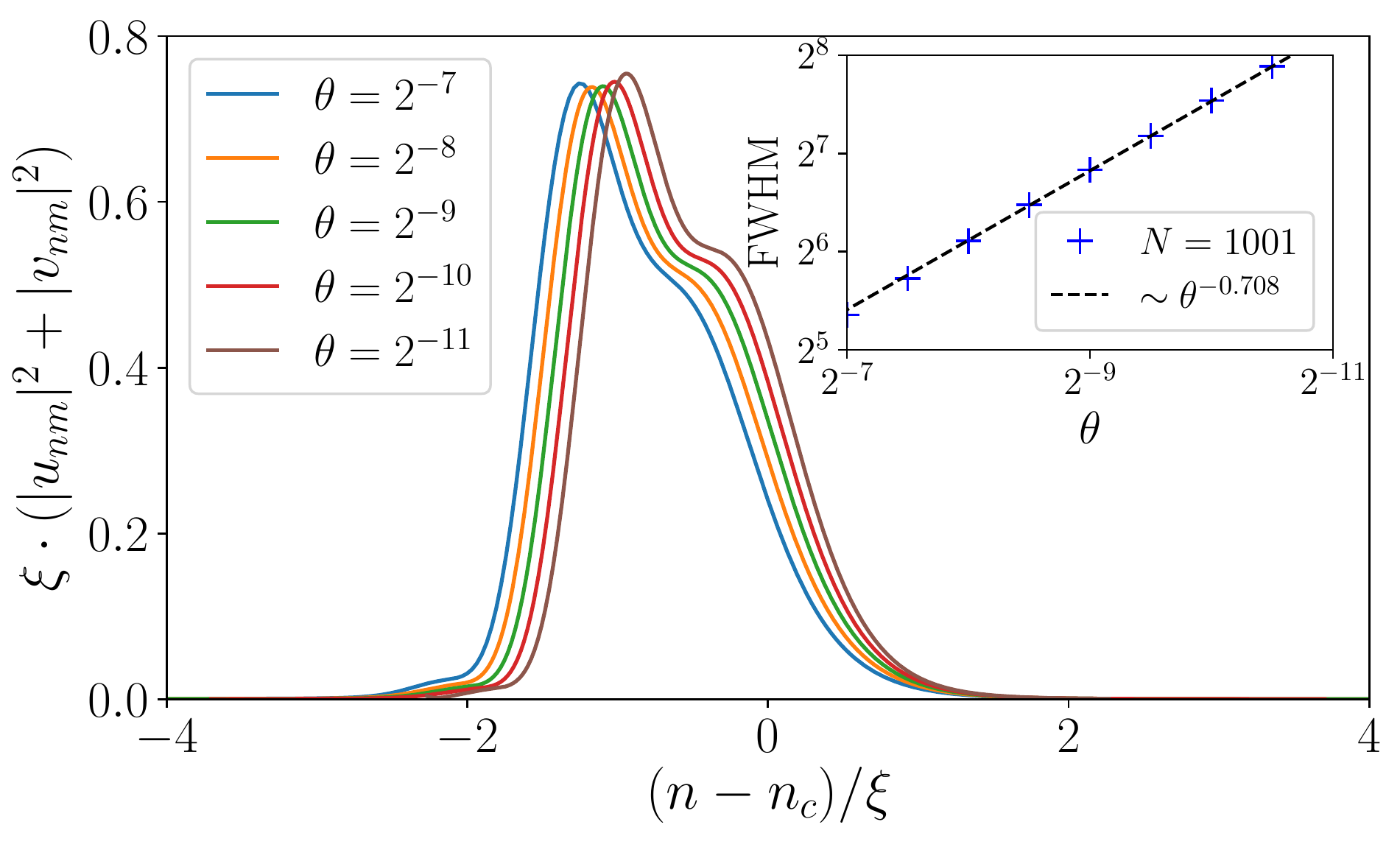}
\caption{
\emph{Localization of the lowest-frequency quasiparticle mode in the long-range extended Ising model.}
The main panel shows scaled density of the first ($m=1$) excited mode, 
$\hat\xi \cdot \left(|u_{nm}|^2+|v_{nm}|^2\right)$,
as a function of the scaled distance from the critical point $(n-n_c)/\hat\xi$
for different slopes $\theta$ in the regime where $1\ll\tilde\xi\ll N$.
Here we set $\tilde{\xi}=\theta^{-0.708}$ in order to obtain the best collapse (up to subleading shift).
This exponent is close to $\hat\xi\propto\theta^{-2/3}$ predicted for $\alpha=3/2$.
In the inset, we show the full width at half maximum (FWHM) of the density distribution as a function of $\theta$ for $N=1001$. The dashed line is the best fit, $\mathrm{ FWHM}=1.37~\theta^{-0.708}$.
}
\label{fig:hatxi}
\end{figure}

In the local density approximation, the healing length would diverge at the critical point at $n_c$, and the local density approximation must break down in the neighborhood of $n_c$. The KZ mechanism in space predicts the size of this neighborhood to be
\be
\tilde{\xi} \propto 
\theta^{-\nu/(1+\nu)} = 
\theta^{-1/\alpha}.
\label{hatxistatic}
\ee
Rather than being non-analytic at $n_c$, the inhomogeneous ground state is rounded off on this static-KZ length scale. The same local density approximation would also predict quasiparticles with zero eigenfrequencies to be localized at $n_c$. As gapless quasiparticles cannot be localized at a point, the KZ mechanism in space predicts that their wave functions are localized within a distance of $\tilde\xi$ around $n_c$, and their eigenfrequencies are finite and proportional to
\be
\tilde\Delta \propto 
\tilde{\xi}^{-z} \propto 
\theta^{z\nu/(1+\nu)} = 
\theta^{(\alpha-1)/\alpha}.
\label{hatDelta}
\ee
This KZ gap sets a scale for the energy of the lowest relevant excitation: $\Delta=\omega_0+\omega_1$. Here, in the bulk, $\omega_0=0$ corresponds to a Majorana mode describing the degeneracy of the two ferromagnetic states to the right of $n_c$ and $\omega_1\propto\tilde\Delta$ is the eigenfrequency of the Bogoliubov quasiparticle localized near~$n_c$. 

The gap $\Delta$ is shown in Fig.~\ref{fig:gapinspace}(a) as a function of slope $\theta$ and system size $N$. There are three regimes of parameters. The most interesting is the central one, where $1\ll\tilde\xi\ll N$, where the KZ equation~\eqref{hatDelta} applies. Indeed, when $N\ll\tilde\xi$, i.e., the case for a ramp with small enough inclination $\theta$, we deal with an effectively homogeneous system at the critical $h_n=h_c$. Consequently, the gap depends on $N$, rather than $\tilde\xi$, as $\Delta\propto N^{-z}$. 
In the opposite extreme limit of $\tilde\xi\ll1$, i.e., when the ramp is effectively a step function, there is a finite non-universal gap that does not depend neither on $N$ nor on $\hat\xi$.

A wave function of the $\omega_1$-quasiparticle is shown in Fig.~\ref{fig:hatxi} for several different slopes in the intermediate regime $1\ll\tilde\xi\ll N$. The collapse demonstrates that the mode is indeed localized near $n_c$ and its localization length is $\tilde\xi$. This localization length explains the crossover to the $N\ll\tilde\xi$ regime, where $N$ limits the quasiparticle size, and the frequency in a finite system at the critical point scales as $\omega_1\propto N^{-z}$.   

The two scales, $\tilde\xi$ and $\tilde\Delta$, can be combined into a velocity scale
\be 
\tilde v = 
\tilde\Delta \tilde\xi \propto 
\theta^{(z-1)\nu/(1+\nu)} =  
\theta^{-(2-\alpha)/\alpha}.
\ee 
As we show below, this velocity discriminate between diabatic and adiabatic inhomogeneous transitions. In contrast to the models with short-range interactions~\cite{inhomo_quantum-a}, for $\alpha<2$ this threshold velocity is a decreasing function of $\theta$. A more shallow front makes the threshold velocity larger.

\subsection{Inhomogeneous transition}
\label{sec:inhomo}

\begin{figure}[t]
\includegraphics[width=0.99\columnwidth]{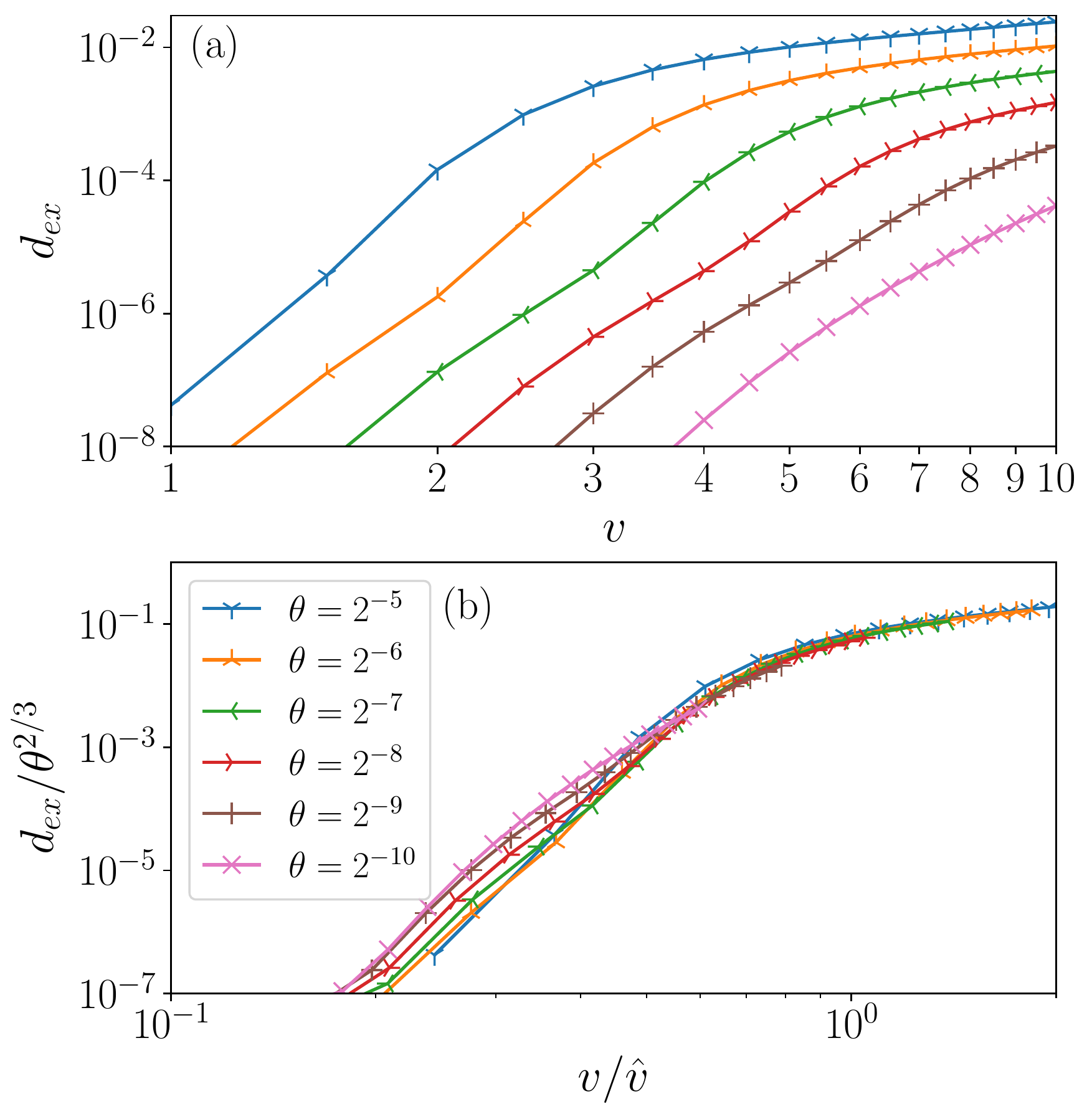}
\caption{
\emph{Density of excitation at the end of the inhomogeneous quench in the long-range extended Ising model.}
Final density of quasiparticle excitations as a function of the front velocity $v$ for different slopes of the front $\theta$. Here $\alpha=3/2$ and {$N=256$}. 
The curves in (a) are rescaled according to Eq.~\eqref{dexinhomoscaledLRE}. 
Here we use $\hat v\propto\hat\Delta\hat\xi$, as in Eq.~\eqref{hatv},
together with our best fits $\tilde\xi\propto\theta^{-0.708}$ and 
$\tilde\Delta\propto\theta^{0.301}$ that yield $\tilde v\propto \theta^{-0.407}$. 
}
\label{fig:dexinhomo}
\end{figure}

Now, we make the ramp in Eq.~\eqref{tanh} to move across the system with velocity $v$,
\begin{align}
h_n(t) = h_c + h_c \tanh \big( \theta (n-vt)/h_c\big),
\label{tanht}
\end{align}
see Fig.~\ref{fig:tanh}. The critical point is moving across the system with velocity $v$. Close to the critical $n_c=v t$ the front can be approximated by a linear ramp $h_n(t)\approx h_{c} + \theta(n-v t)$. We start with an initial $t_{i} = -C h_{c}/\theta v$ and stop at final $t_{f} = N/v+C h_{c}/\theta v$.  The ramp, when watched locally at a fixed $n$, looks like a homogeneous quench with a quench time $\tau_Q$ satisfying
\be 
v \theta = \frac{1}{\tau_Q}.
\label{tauQinhomo}
\ee 
Figure~\ref{fig:dexinhomo} shows the final density of the excited Bogoliubov quasiparticles after the ramp passes the whole chain. The result depends on velocity $v$ and the slope $\theta$. For any slope, when $v$ is faster than $\tilde v$ in Eq.~\eqref{tildev}, the transition should be effectively homogeneous because the finite gap $\tilde\Delta$ is too small to make it adiabatic. In this regime, a combination of Eqs.~\eqref{dexhomo} and \eqref{tauQinhomo} yields
\be
d_{\rm ex}\propto 
\left(v\theta\right)^{\nu/(1+z\nu)}.
\label{dexinhomo}
\ee 
It can be conveniently rewritten as a power law \footnote{Subleading corrections to the power-law scaling can be expected, though as long as they appear in the form of $v/\hat v$, as for transverse-field Ising model~\cite{inhomo_quantum-a}, they do not violate the collapse upon rescaling.}
\be 
 d_{\rm ex}\theta^{-\nu/(1+\nu)}\propto 
\left(\frac{v}{\tilde v}\right)^{\nu/(1+z\nu)},
\label{dexinhomoscaled}
\ee 
that for long-range extended Ising model reduces to
\be 
d_{\rm ex} \theta^{-1/\alpha} \propto \left(\frac{v}{\tilde v}\right)^{1/2(\alpha-1)}.
\label{dexinhomoscaledLRE}
\ee 
This power law---valid in the effectively homogeneous KZ regime---motivates the scaled plot in Fig.~\ref{fig:dexinhomo} of $d_{\rm ex}/\theta^{1/\alpha}$ as a function of $v/\tilde v$. This scaling collapses the curves for different slopes $\theta$. The collapse demonstrates that indeed, $\tilde v$ is the threshold velocity that discriminates between an effectively homogeneous transition for $v\gg\tilde v$ and an adiabatic one for $v\ll\tilde v$. This conclusion is valid in the central regime, $1\ll\hat\xi\ll N$, when the ramp is neither too shallow nor too steep.

\section{Long-range Ising chain } 
\label{sec:LRI}

The long-range Ising chain reads:
\be
H=
-\sum_{n=1}^N
\left(
 h_n \sigma^z_n +
\sum_{r=1}^{N-n} 
J_r\sigma^x_n\sigma^x_{n+r}
\right). \label{Hamil_LRI}
\ee 
Here $J_r = \frac{1}{r^\alpha}$ is the coupling strength between two spins separated by a distance $r$, and $h_{n}$ is the transverse field. Unlike its extended form described in the previous section, the Hamiltonian~\eqref{Hamil_LRI} cannot be solved by mapping onto the free-fermionic model. For that reason, even the most basic equilibrium results, including the location of the critical point and values of the critical exponents $z$ and $\nu$, require sophisticated numerical methods such as tensor network techniques~\cite{orus_review_2014}.
The latter have recently been used in Refs.~\onlinecite{KZLR2, KZLR3} to obtain the data on the critical points and show the validity of Kibble-Zurek physics in long-range Ising models. 

\begin{figure}[t]
\includegraphics[width=0.99\columnwidth]{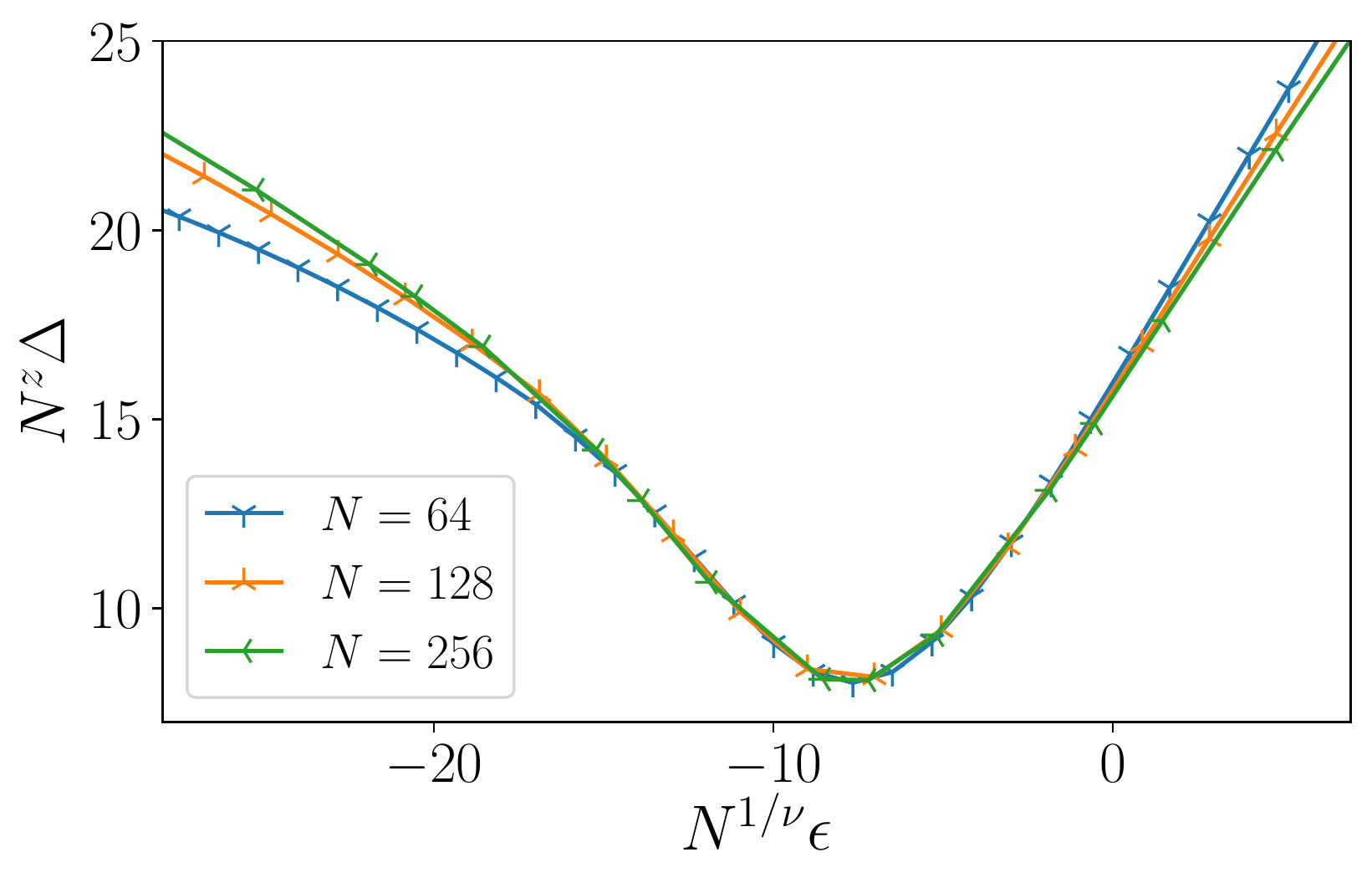}
\caption{{\it Rescaled energy gap as a function of the distance from the critical point in the long-range Ising chain}. Results for the homogeneous $h_n = h_c -\epsilon$, where the gap is calculated after projecting the Hamiltonian onto the subspace with even parity ${\mathcal P}$, see Eq.~\eqref{eq:parity}. The collapse around the minimum coraborates the obtained values of the exponents $\nu \approx 1.3$ and $z \approx 0.48$.}
\label{fig:energy_gap_alpha2}
\end{figure}

There are two important limits to this model that can be exactly solved: for $\alpha \rightarrow \infty$, we retrieve the standard short-range transverse Ising chain, and for $\alpha = 0$, we recover the Lipkin-Meshkov-Glick model with infinite-range couplings. While the limiting cases have exact solutions, the intermediate-range has to be solved numerically. Several studies~\cite{Fisher1972,Fey2016,Defenu2017,Zhu2018} show that three broad regimes exist depending on the value of the coupling exponent $\alpha$: the mean-field universality class regime for $1<\alpha < 5/3$, the continuously varying universality class regime for $5/3 < \alpha < 3$, and the Ising universality class regime for $\alpha > 3$.  In the mean-field regime, the correlation length critical exponent $\nu = 1/(\alpha-1)$ and the dynamical critical exponent $z= (\alpha -1)/2$ are given by mean-field values. In the Ising regime, the critical exponents $\nu = 1$ and $z=1$ are also well-known. However, in the intermediate regime with $5/3< \alpha <3$, the critical exponents $\nu$ and $z$ change monotonously with no analytical values known. In the following, we will numerically determine the critical exponents $\nu$ and $z$ for a particular $\alpha$ in this intermediate regime and use it to analyze the consequences of an inhomogeneous driving of the transverse field. On the one hand, long-range interaction effects are appreciable only for $\alpha < 2.25$~\cite{Koffel2012}. On the other hand, finite-size effects are becoming strong for smaller values of $\alpha$, strongly limiting the possibility of effective simulations.  These, compounded with aesthetic reasons, prompts us to focus on  $\alpha = 2$ for further calculations. For completeness, we have also tested $\alpha=2.2$, arriving at the same conclusions.

\begin{figure}[t]
\includegraphics[width=0.99\columnwidth]{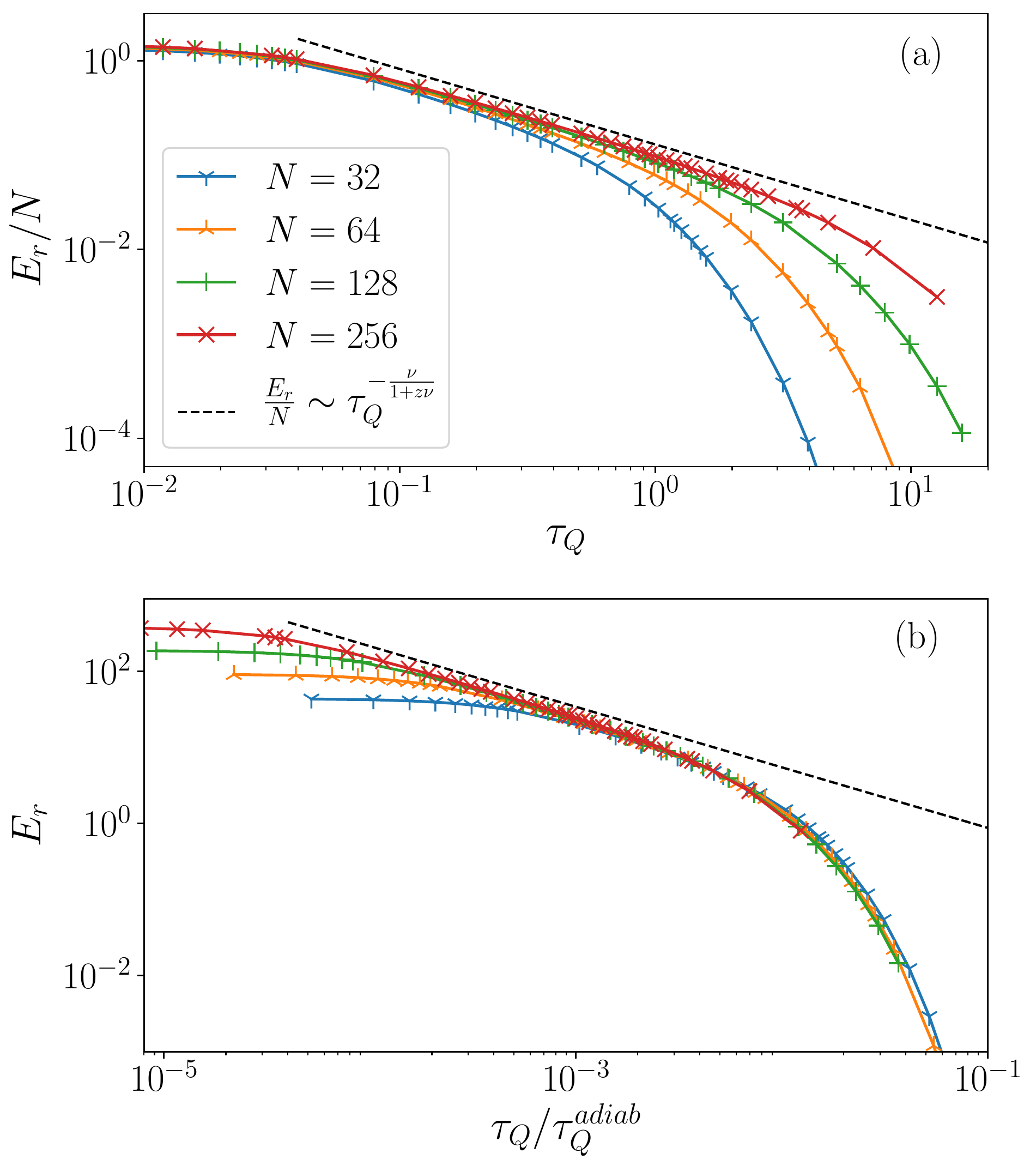}
\caption{{\it Residual energy per spin $E_{r}/N$ at the end of the homogeneous quench in the long-range Ising model}. The ramp follows Eq.~\eqref{homoramp}, crossing the critical point at quench time $\tau_Q$. The long-range interactions, $J_r$, vanish with exponent $\alpha=2$. In (a), $E_{r}/N$ decays like $\tau_Q^{-\nu/(z\nu+1)}$ for faster quenches as predicted by Eq.~\eqref{res_energy_formula}. Finite-size effects become visible in the limit of larger $\tau_Q$. We show their universal character in (b), where we present rescaled density $(E_{r}/N)\times N$ as a function of quench time $\tau_{Q}$ rescaled by $\tau_{Q}^{\rm adiab} \propto N^{\frac{1+z\nu}{\nu}}=N^{1.25}$. The collapse of the plots for different $\tau_Q$ demonstrates the crossover from the KZ to the adiabatic regime predicted by Eq.~\eqref{homoadiab}.}
\label{fig:homo_alpha2}
\end{figure}

In the homogeneous case, we set the transverse field, $h_{n} = h_{c}-\epsilon$, with $\epsilon>0$ in the ferromagnetic phase and $\epsilon<0$ is the paramagnetic case. To determine the value of $h_c$ for $\alpha = 2$, we perform a finite-size scaling of the position of the minimum of the energy gap (in the finite system) between the ground state and the first excited state [parity operator ${\mathcal P}$ in Eq.~\eqref{eq:parity} is also a good quantum number  for a long-range Ising model, so we first project the Hamiltonian onto the subspace with even parity].
We obtain $h_{c} \approx 2.528$, and the correlation length exponent $\nu \approx 1.3$. Subsequently, at the obtained $h_c$, we perform a finite-size scaling analysis of the gap to find the dynamical exponent $z \approx 0.48$, see  the Appendix~\ref{critical_pts_exponents}. To further corroborate the above results, in Fig.~\ref{fig:energy_gap_alpha2}, we show the collapse of the rescaled energy gaps near the critical point for different system sizes.

In the following, we simulate homogeneous and inhomogeneous quenches through a critical point. The time evolution is performed using the time-dependent variational principle (TDVP)~\cite{Haegeman2011, Haegeman2016} \footnote{In particular, we use a combination of the $1$-site and $2$-sites TDVP updates, that combines the time efficiency of the $1$-site scheme with the possibility to locally enlarge the matrix product state bond dimension when necessary through the $2$-sites gates, see, e.g., the supplementary material of Ref.~\onlinecite{Rams2020} for more details. Such an approach is well suited to address the inhomogeneous nature of the problem at hand. The presented results have been obtained with a dynamically grown bond dimension up to $\chi_{\rm max}= 150$ (i.e., a parameter which ultimately controls the convergence), which we have further checked against higher $\chi_{\rm max}$ for selected points.} for matrix product states, which allows one to handle any Hamiltonian expressed as a matrix product operator---including the one in Eq.~\eqref{Hamil_LRI}.  The ground states and first-excited states have been obtained using standard density-matrix renormalization group algorithms~\cite{schollwock_review_2011}.

\subsection{Homogeneous transition}
\label{sec:KZhomoI}
In this section, we perform a slow homogeneous quench of the transverse field from the paramagnetic phase to the ferromagnetic phase. We choose the same ramp as given in Eq.~\eqref{homoramp}, between the initial time $t_{i} =-C h_{c}\tau_{Q}$ and the final time $t_{f} = C h_{c} \tau_Q$. We choose $C = 4$, which ensures we start and end quite deep in either phase.

To quantify the excitations induced by a quench, here we employ the residual energy,
\be
E_{r} = E_{f} -E_{gs},
\label{res_energy}
\ee 
where $E_{gs}$ is the ground state energy of $H(t_{f})$ and $E_{f} = 
\braket{\Psi(t_{f})|H(t_{f})|\Psi(t_{f})}$ is the system's energy after the quench. The expected KZ scaling prediction for slow quench (and in the thermodynamic limit) reads
\be
E_{r}/N \sim \tau_{Q}^{-\nu/(z\nu+1)}.
\label{res_energy_formula}
\ee

In Fig.~\ref{fig:homo_alpha2}(a), we plot the residual energy per site $E_{r}/N$ for $N = 32, 64, 128$ and $256$. Three regions can be clearly visible: (a) for fast quenches $E_{r}/N$ is at its maximum and does not show apparent dependence on size $N$; (b) the intermediate region follows the quantum Kibble-Zurek scaling of $\tau_{Q}^{-0.80}$ with $z \approx 0.48$ and $\nu \approx 1.3$; and (c) the exponential decay is expected upon increasing $\tau_{Q}$ further (we cannot extract this due to the limited accuracy of our simulations).
The transition between (b) and (c) is expected when $\hat{\xi} \sim N$. Indeed upon rescaling $\tau_{Q}$ with $\tau_{Q}^{\rm adiab}$ [see Eq.~\eqref{homoadiab}], and simultaneously the energy density consistently with the scaling in Eq.~\eqref{res_energy_formula} [giving $N E_{r}/N = E_{r}$], the curves for different system sizes collapse in Fig.~\ref{fig:homo_alpha2}(b). This signifies the crossover from the KZ regime to the adiabatic regime.

\begin{figure}[t!]
\includegraphics[width=0.99\columnwidth]{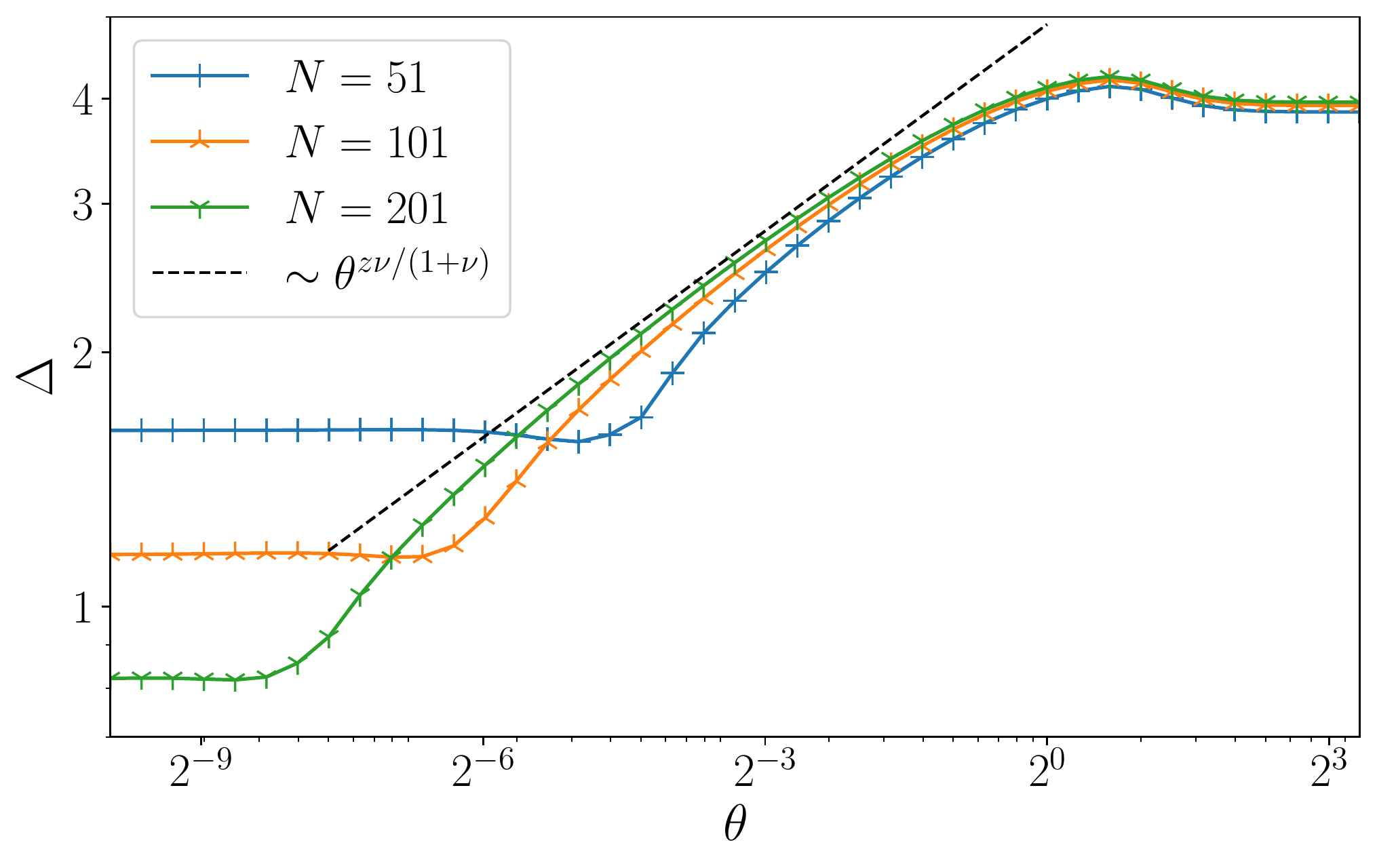}
\caption{\textit{Energy gap $\Delta$ as a function of inclination parameter of the inhomogeneous front $\theta$ for different system sizes $N$ in the long-range Ising chain}. The external field follows Eq.~\eqref{tanht} and the gap is calculated in the even-parity subspace. As predicted by Eq.~\eqref{hatDelta_alpha2}, the central part is consistent with the KZM-in-space prediction $\Delta \propto \theta^{\frac{z\nu}{1+\nu}} \approx \theta^{0.27}$.}
\label{fig:kzm_in_space_alpha_2}
\end{figure}

The physical reason for these different regimes is the participation weight of the quenched state among the different energy levels.  For the fast quenches, the dynamics take place among a vast proportion of excited energy levels. For the intermediate dynamics, only the low-lying excited states participate, allowing one to observe universal behavior. The exponential decay for the largest $\tau_{Q}$'s reflects the Landau-Zener formula, which denotes the probability for a particle in the ground state to jump into the first excited state as a function of the driving timescale $\tau_{Q}$ in a single avoided-level crossing.

\subsection{Transition in space}

Similarly, as for the extended case, here we consider the static inhomogeneous field given by Eq.~\eqref{tanh}. 
In Fig.~\ref{fig:kzm_in_space_alpha_2}, we plot the energy gap $\Delta$ between the ground state and first excited state of the Hamiltonian (after projecting the latter onto an even parity subspace) for system sizes $N = 51, 101$ and $201$. Following the analysis in the extended Ising model, we identify the extent of the intermediate region. This instructs us on choosing the inclination range for further simulations of the inhomogeneous driving. The KZ prediction for the dependence of the energy gap on the inclination $\theta$ in the intermediate region is
\be
\hat\Delta \propto 
\theta^{z\nu/(1+\nu)} = \theta^{0.27}.
\label{hatDelta_alpha2}
\ee
As can be seen in Fig.~\ref{fig:kzm_in_space_alpha_2}, it is satisfied to an appreciable extent.

\begin{figure}[t!]
\includegraphics[width=0.99\columnwidth]{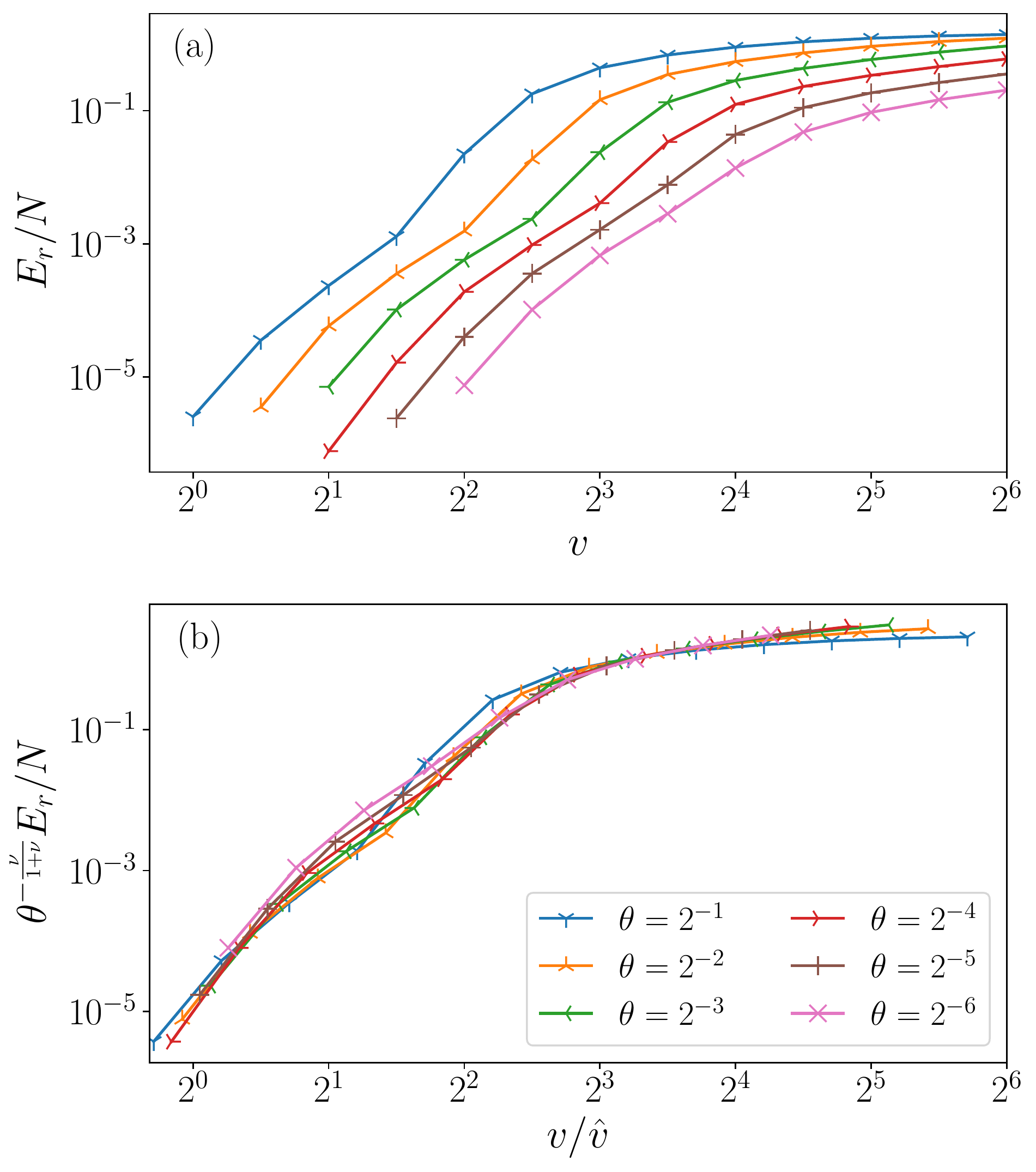}
    \caption{{\it Residual energy per spin $E_{r}/N$ at the end of the inhomogeneous quench in the long-range Ising model}. The ramp follows Eq.~\eqref{tanht}, crossing the critical point at an inclination $\theta$ and velocity $v$ of the inhomogeneous front. The long-range interactions, $J_r$, vanish with exponent $\alpha=2$. In (a), $E_{r}/N$ is plotted for several values of $\theta = 2^{-1},2^{-2},...,2^{-6}$. The plots are shown in log-log scale. We show their universal character in (b), where we present rescaled density $ \theta^{-\frac{\nu}{1+\nu}} E_{r}/N$ as a function of velocity $v$ scaled by $\tilde{v} \sim \theta^{(z-1)\nu/(1+\nu)}\approx \theta^{-0.29} $. The collapse of the plots for different $\theta$'s demonstrates the existence of a characteristic crossover velocity $\tilde{v}$ which separates the diabatic homogeneous regime and the adiabatic inhomogeneous regime.}
\label{fig:inhomo_LRI}
\end{figure}

\subsection{Inhomogeneous transition}
\label{sec:inhomoLRI}

Finally, we use the same profile, see Eq.~\eqref{tanht}, to drive an inhomogeneous quench across the system.  As discussed in Sec.~\ref{sec:inhomo}, a single lattice site $n$ effectively undergoes a local change of transverse field from $h_{n}\approx 2h_{c}$ to $h_{n}\approx 0$ (with the actual moment of the transition delayed from site to site). In Fig.~\ref{fig:inhomo_LRI}(a), we show the final residual energy as a function of velocity $v$ for different values of $\theta$. The latter has been chosen by keeping in mind the extent of the intermediate region in Fig.~\ref{fig:kzm_in_space_alpha_2}. The residual energy vanishes quickly below the characteristic $\tilde{v}$. 

To quantify the universal behavior, we proceed similarly as in the previous section. Equations \eqref{hatv} and \eqref{dexinhomo} provide us with two effective scales: one for velocity $\tilde v \sim \theta^{(z-1)\nu/(1+\nu)}$ and one for the residual energy $E_{r}/N \sim \theta^{\frac{\nu}{1+\nu}}$. 
Upon rescaling according to these formulas, see Fig.~\ref{fig:inhomo_LRI}(b), curves obtained for different slopes $\theta$ collapse. The collapse corroborates the existence of a characteristic KZ velocity scale, $\tilde{v}$. A velocity $v \gg \tilde{v}$ would result in an effectively homogeneous transition, while $v \ll \tilde{v}$ would lead to an adiabatic transition.

\section{Shortcut to adiabaticity}
\label{sec:shortcuts}

The minimal time needed for the ramp to cross the system adiabatically, for a given inclination $\theta$, is 
\be 
T=\frac{N}{\tilde v}+\frac{A}{\theta \tilde v}. 
\label{T}
\ee 
Here $N/\tilde v$ is the time needed for the critical point to cross the chain of length $N$ with velocity $\tilde v$. The second term, with $A={\cal O}(1)$, is an overhead needed to begin/end the ramp with the whole chain deep in the symmetric/symmetry-broken phase. For instance, when we evolve the right-moving tanh-ramp in Eq.~\eqref{tanht} from a initial position (such that $h=1+\epsilon_0$ on the left end of the chain) to a final position (such that $h=1-\epsilon_0$ on the right end of the chain), then $A=2\tanh^{-1}(\epsilon_0)$. Minimization of $T$ with respect to $\theta$, for $z<1$, yields the optimal slope
\be 
\theta_{\rm opt} = A \frac{1+z\nu}{(1-z)\nu} N^{-1}.
\ee 
For this slope, the optimal time is 
\be 
T_{\rm opt} \propto N^{(1+z\nu)/(1+\nu)}.
\ee 
It is the shortest time that allows for adiabatic evolution with an inhomogeneous ramp moving at a constant velocity of $\tilde v$. 
We compare this time with the optimal time for a homogeneous quench
to see if an inhomogeneous quench allows for any advantage. The homogeneous transition becomes adiabatic for quench times such that KZ correlation length $\hat\xi$ becomes comparable with the system size $N$. Consequently, the minimal adiabatic quench time is
\be 
\tau_Q^{\rm adiab} \propto N^{(1+z\nu)/\nu}. 
\label{xxz}
\ee 
The ratio of the optimal inhomogeneous to the optimal homogeneous time is
\be 
\frac{T_{\rm opt}}{\tau_Q^{\rm adiab}}=N^{-(1+z\nu)/\nu(1+\nu)}.
\ee 
As we can see, the optimal inhomogeneous transition is always advantageous over the optimal homogeneous quench, which we further illustrate below with our two examples. In this analysis, we limit ourselves to consider only homogeneous quenches that have constant $\tau_Q$. Having inhomogeneous-in-time quench---adjusting the quench rate to the instantaneous size of the gap, e.g., following a power-law quench~\cite{QKZteor-h}---would give further opportunities for homogeneous quenches. However, this would require additional precise knowledge regarding the exact position of the minimum and the behavior of the gap around it, as well as precise control. Here, we focus on comparing more robust strategies, where such detailed knowledge is not needed.

\subsection{Long-range extended Ising chain}

The optimal slope satisfies
\be 
N \theta_{\rm opt} = 2A\frac{(\alpha-1)}{(2-\alpha)}.
\ee 
In the limit of effectively local interaction (when $\alpha\to 2^-$), the optimal ramp becomes very steep compared to the chain's length. That is consistent with the results for the local transverse-field Ising chain where a steep ramp quenching a few spins at a time was found to be the optimal one---the many-body critical (and universal) nature of the problem is lost in the limit of one-spin-at-a-time quench, i.e., $\theta \to 1$, and this type of driving turns out to be sub-optimal~\cite{inhomo_quantum-d}. In the opposite, non-local limit, when $\alpha\to 1^+$, the optimal ramp becomes virtually homogeneous compared to the chain's length. In any case, the optimal time is
\be 
T_{\rm opt} \propto N^{2(\alpha-1)/\alpha}.
\ee 
This is to be compared with the optimal homogeneous quench time
\be 
\tau_Q^{\rm adiab} \propto N^{(1+z\nu)/\nu} {=N^{2(\alpha - 1)}}. 
\ee 
Their ratio is
\be 
\frac{T_{\rm opt}}{\tau_Q^{\rm adiab}} \propto N^{-2(\alpha-1)^2/\alpha}.
\ee 
For $1<\alpha<2$, it is always advantageous to use the optimal inhomogeneous quench over the homogeneous one. In the local limit, $\alpha\to2^-$, the ratio becomes proportional to $1/N$. In the non-local limit, $\alpha\to 1^+$, it tends to $O(1)$. The inhomogeneous transition has no advantage when one increases the non-locality of interactions in which case, in fact, there is little difference between the homogeneous and inhomogeneous transition. 

\subsection{Long-range Ising chain}

When $\alpha>3$ we end up in the Ising universality class where, just as in the local limit of the long-range extended Ising chain, the inhomogenous advantage $T_{\rm opt}/\tau_Q^{\rm adiab}\propto 1/N$ is achieved for a steep ramp where spins are quenched a few at a time. On the other hand, when $1<\alpha<5/3$, in the mean-field universality class regime with $\nu=(\alpha-1)^{-1}$ and $z=\frac{\alpha-1}{2}$, $T_{\rm opt}/\tau_Q^{\rm adiab} \propto N^{-3(\alpha -1)^2/2\alpha} $ and the inhomogeneous quench shows a clear advantage over the homogeneous one. However, as in the extended long-range Ising model, in the non-local limit, when $\alpha\to1^+$, the advantage disappears. When all pairs of spins are coupled with equal strength, making the position of a spin along the chain a mere label, then there is no difference between the homogeneous and inhomogeneous transitions.  

In the continuously varying universality class regime for $5/3<\alpha<3$, we considered a representative $\alpha=2$. With the numerically estimated critical exponents, we obtain the optimal slope,
\be 
N \theta_{\rm opt} = 2.40 A,
\ee 
which allows for the optimal inhomogeneous-driving time
\be 
T_{\rm opt} \propto N^{0.70}.
\ee 
Comparing with the minimal adiabatic homogeneous quench time, we get
\be 
\frac{T_{\rm opt}}{\tau_Q^{\rm adiab}} \propto N^{-0.54}.
\ee
Therefore, carefully optimized inhomogeneous quench provides clear advantage as a shortcut to adiabaticity.

\section{Summary}
\label{sec:summary}

We demonstrated a clear crossover between the adiabatic regime of an inhomogeneous transition and the Kibble-Zurek regime, where the transition proceeds like in a uniform system. The latter regime persists despite the non-local nature of the interaction. In principle, irrespective of the speed of the critical front, the interaction might be able to instantaneously communicate  the initial choice of symmetry breaking behind the critical front to the spins ahead of the front and bias them to make the same choice when crossing the critical point. As a consequence, excitations could have been expected to be suppressed in the symmetry-broken phase behind the front. However, this instantaneous bias becomes irrelevant when the critical front's spatial velocity starts exceeding $\tilde v$---a characteristic speed derived from the static inhomogeneous KZ mechanism.  We support this observation with numerical studies in two long-range models, one integrable and one non-integrable, corroborating its universal character.

Even though $\tilde v$ is not an upper, but a lower speed limit of excitations, for a critical front slower than $\tilde v$, the inhomogeneous transition becomes adiabatic in a similar way as for local interactions where the finite speed of excitations limits the communication across the critical front. The slope of the critical front can be optimized to achieve an adiabatic transition in timescales that are shorter than the shortest time of the adiabatic transition achievable by a uniform protocol.

\acknowledgements
AS would like to thank Titas Chanda and Anna Francuz for useful discussions on some technical aspects of the work. We would also like to thank Daniel Jaschke for providing data for Fig.\ $2$ of Ref.~\onlinecite{KZLR2}. Part of the numerical calculations were performed in MATLAB with the help of the \texttt{ncon} function\cite{ncon} for tensor contractions. We acknowledge funding by the National Science Centre (NCN), Poland under projects No.~2016/23/B/ST3/00830 (JD) and No.~2016/23/D/ST3/00384 (MMR), 
and NCN together with European Union through QuantERA ERA NET Program No.~2017/25/Z/ST2/03028 (AS,DS).
This research was carried out with the equipment purchased thanks to the financial support of the European Regional Development Fund in the framework of the Polish Innovation Economy Operational Program (Contract No. POIG.02.01.00-12-023/08). 

\appendix

\section{Obtaining critical point and critical exponents of the long-range Ising Model}
\label{critical_pts_exponents}

In the main text, we have already emphasized that the analytical determination of critical points and exponents is not feasible in a long-range Ising Hamiltonian.  Here, we resort to density matrix renormalization group techniques~\cite{schollwock_review_2011} using matrix product states to calculate the energy gap between the ground state and the first relevant excited state. The location of the minimum of the gap in the thermodynamic limit $N \rightarrow \infty$ (and for infinite bond dimension $\chi$) would return the exact location of the critical point. Having inhomogeneous systems in mind, we focus on finite systems performing finite-size scaling (and with maximal bond dimension $\chi_{\rm max} = 200$ which is large enough for the required precision). In particular, we look at the location of the minimal gap, $h_{c}(N)$, for the given system size $N$.
It is expected that 
\be
h_{c}(N) = \lim_{N \rightarrow \infty} h_{c}(N) + cN^{-1/\nu}.
\label{cardy_formula}
\ee
\begin{figure}[t!]
\includegraphics[width=0.99\columnwidth]{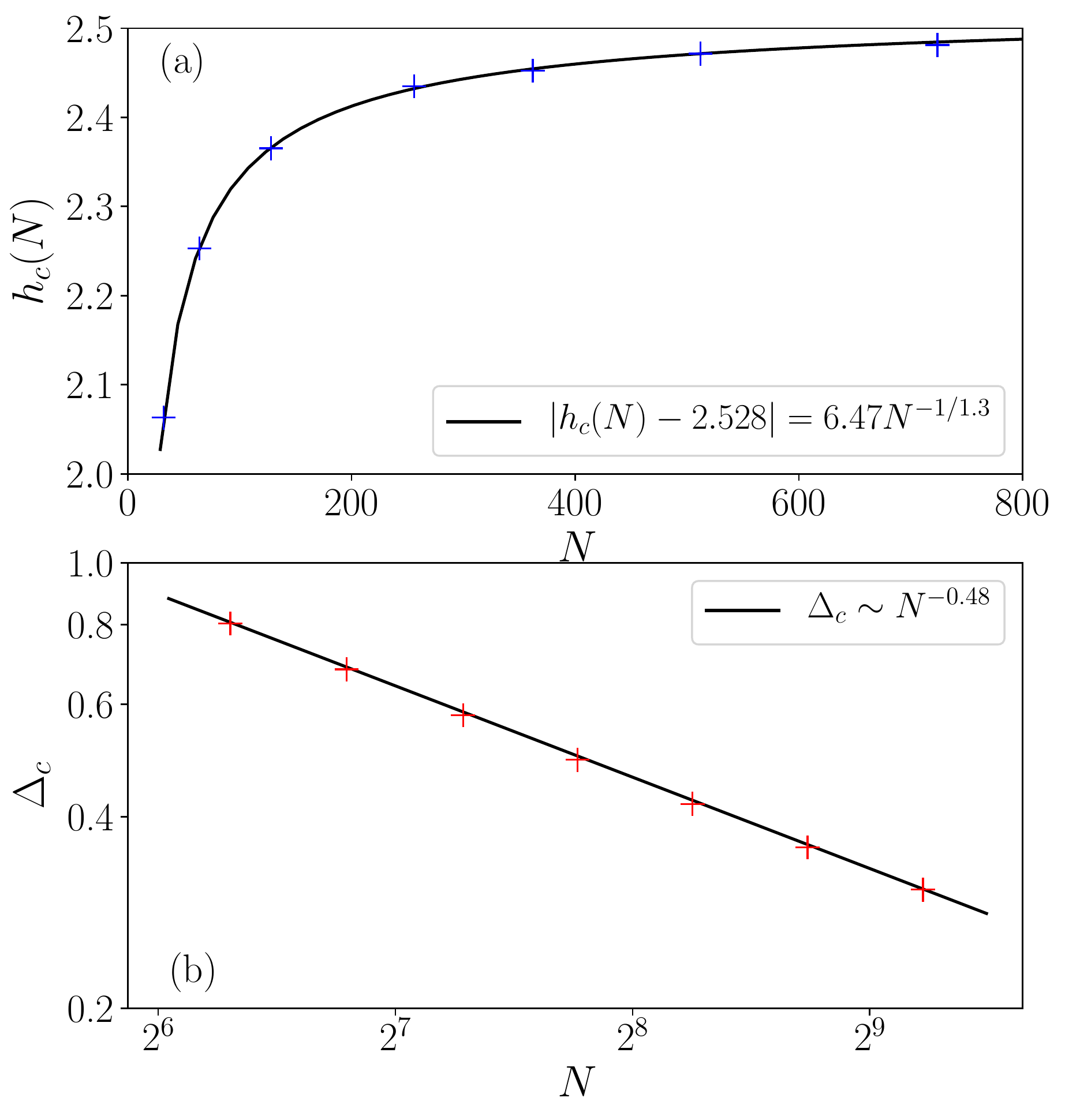}
\caption{(a) \textit{Location of the minimal gap, $h_{c}(N)$, as a function of the system size $N$ in the long-range Ising chain}. The fit give an estimate of the critical point $h_{c} \approx 2.528$ and the correlation length exponent $\nu \approx 1.3$ for coupling $J_r$ following a power law with exponent $\alpha=2$. (b) The finite-size scaling of the energy gap $\Delta_{c}$ calculated at the critical point $h_{c}$. The slope of the best fit gives an estimate of the dynamical exponent $z \approx 0.48$.}
\label{fig:min_gap_alphaz}
\end{figure}
To prevent the over-representation of smaller system sizes, during fitting, we add a weight proportional to $N$ to each point, following the strategy by Jaschke {\it et. al.}~\cite{KZLR2}. The non-linear fit predicts $h_{c} \approx 2.528$ and $\nu \approx 1.3$. Next, we determine the dynamic exponent $z$ through finite-size scaling of the energy gap at the critical point, see Fig.~\ref{fig:min_gap_alphaz}(b) and obtain $z \approx 0.48$.
%
\bibliographystyle{apsrev4-1}
\bibliography{main.bbl}
%
\end{document}